\documentclass[11pt,draftclsnofoot,onecolumn]{IEEEtran}
\IEEEoverridecommandlockouts                              	
\title{Asymptotic Stability and Decay Rates of Homogeneous Positive Systems With\\ Bounded and Unbounded Delays}

\author{Hamid Reza Feyzmahdavian$^{\star}$, Themistoklis Charalambous$^{\star}$, and Mikael Johansson$^{\star}$
\thanks{$^{\star}$Department of Automatic Control, School of Electrical Engineering and ACCESS Linnaeus Center, Royal Institute of Technology (KTH), SE-100 44 Stockholm, Sweden.
Emails: {\tt \{hamidrez, themisc, mikaelj\}@kth.se.}}
}

\usepackage[T1]{fontenc}
\usepackage{graphicx}
\graphicspath{{figures/}}
\usepackage{amssymb,amsmath,amsfonts}
\usepackage{bm}
\usepackage{epsfig}
\usepackage{epstopdf}
\usepackage{verbatim}
\usepackage[usenames,dvipsnames]{color}
\usepackage{flushend}
\usepackage{cite}

\newtheorem{theorem}{\normalfont \bfseries Theorem}
\let\oldtheorem\theorem
\renewcommand{\theorem}{\oldtheorem\em}

\newtheorem{definition}{\normalfont \bfseries Definition}
\let\olddefinition\definition
\renewcommand{\definition}{\olddefinition\em}

\newtheorem{proposition}{\normalfont \bfseries Proposition}
\let\oldproposition\proposition
\renewcommand{\proposition}{\oldproposition\em}

\newtheorem{corollary}{\normalfont \bfseries Corollary}
\let\oldcorollary\corollary
\renewcommand{\corollary}{\oldcorollary\em}

\newtheorem{example}{\normalfont \bfseries Example}
\let\oldexample\example
\renewcommand{\example}{\oldexample\em}

\newtheorem{remark}{\normalfont \bfseries Remark}
\let\oldremark\remark
\renewcommand{\remark}{\oldremark\em}

\newtheorem{assumption}{\normalfont \bfseries Assumption}
\let\oldassumption\assumption
\renewcommand{\assumption}{\oldassumption\em}

\def\bnull{\bm{0}}
\def\r{\bm{r}}
\def\x{\bm{x}}
\def\y{\bm{y}}
\def\v{\bm{v}}
\def\f{\bm{f}}
\def\g{\bm{g}}
\def\bphi{\bm{\varphi}}

\begin{document}

\maketitle

%
%

\begin{abstract}
There are several results on the stability of nonlinear positive systems in the presence of time delays. However, most of them assume that the delays are constant. This paper considers {\em time-varying}, possibly unbounded, delays and establishes asymptotic stability and bounds the decay rate of a significant class of nonlinear positive systems which includes positive linear systems as a special case. Specifically, we present a necessary and sufficient condition for delay-independent stability of continuous-time positive systems whose vector fields are cooperative and homogeneous. We show that global asymptotic stability of such systems is independent of the magnitude and variation of the time delays. For various classes of time delays, we are able to derive explicit expressions that quantify the decay rates of positive systems. We also provide the corresponding counterparts for discrete-time positive systems whose vector fields are non-decreasing and homogeneous.
\end{abstract}

%
%

\section{Introduction}\label{sec:intro}

Many real-world processes in areas such as economics, biology, ecology and communications deal with physical quantities that cannot attain negative values. The state trajectories of dynamical models characterizing such processes should thus be constrained to stay within the positive orthant. Such systems are commonly referred to as {\em positive systems} \cite{Smith:95,Farina:00,Haddad:10}. Due to their importance and broad applications, a large body of literature has been concerned with the analysis and control of positive systems (see, e.g., \cite{Jacquez:93,Valcher:97,Van:98,Leenheer:01,Aeyels:02,Dashkovski:06,Rami:07,Mason:07,Ngoc:09,Fainshil:09,2009:Florian,Ruffer:10,Rantzer:11,Tanaka:11,Fornasini:11,
li:11,Ebihara:13,Feyzmahdavian:13-0,Briat:13} and references therein).

In distributed systems where exchange of information is involved, delays are inevitable. For this reason,  a considerable effort has been devoted to characterizing the stability and performance of systems with delays (see, e.g., \cite{Hale:93,Fridman:02,Wu:04,Gao:07,He:07,Peet:09,Sipahi:11} and references therein). Recently, the stability of delayed positive linear systems has received significant attention \cite{Ngoc:06,Liu:09,Rami:09,Liu:10,Haddad:04} and it has been shown that such systems are insensitive to certain classes of time delays, in the sense that a positive linear system with time delays is asymptotically stable if the corresponding delay-free system is asymptotically stable. This is a surprising property, since the stability of general dynamical systems typically depends on the magnitude and variation of the time delays.

While the asymptotic stability of positive linear systems in the presence of time delays has been thoroughly investigated, the theory for {\em nonlinear} positive systems is considerably less well-developed (see, e.g., \cite{Haddad:04,Mason:09,Vahid:10} for exceptions). In particular, \cite{Mason:09} showed that the asymptotic stability of a particular class of nonlinear positive systems whose vector fields are {\em cooperative} and {\em homogenous} of degree zero does not depend on the magnitude of {\em constant} delays. A similar result for cooperative systems that are homogeneous of any degree was given in \cite{Vahid:10}, also under the assumption of constant delays. Extensions of these results to {\em time-varying} delays are, however, not trivial. The main reason for this is that the proof technique in \cite{Mason:09,Vahid:10} relies on a fundamental monotonicity property of trajectories of cooperative systems, which does not hold when the delays are time-varying. To the best of our knowledge, there have been rather few studies on stability of nonlinear positive systems with time-varying delays (see, e.g., \cite{Feyzmahdavian:14,Ngoc:13,Feyzmahdavian:14-1}).

At this point, it is worth noting that the results for positive linear systems cited above consider {\em bounded} delays. However, in some cases, it is not possible to a priori guarantee that the delays will be bounded, but the state evolution might be affected by the entire history of states. It is then natural to ask if the insensitivity properties of positive linear systems with respect to time delays will hold also for {\em unbounded} delays. In \cite{Liu:2011}, it was shown that, for a particular class of unbounded delays, this is indeed the case. Extensions of this result to more general classes of unbounded delays were given in \cite{Sun:12,Feyzmahdavian:13} for continuous- and discrete-time positive linear systems, respectively. However, \cite{Liu:2011,Sun:12,Feyzmahdavian:13} did not quantify how various bounds on the delay evolution impact the decay rate of positive linear systems.

This paper establishes delay-independent stability of a class of nonlinear positive systems, which includes positive linear systems as a special case, and allows for time-varying, possibly unbounded, delays. The proof technique, which uses neither the Lyapunov-Krasovskii functional method widely used to analyse positive systems with constant delays \cite{Haddad:04} nor the approach used in \cite{Mason:09,Vahid:10}, allows us to impose minimal restrictions on the delays. Specifically, we make the following contributions:
\begin{itemize}
\item We derive a set of necessary and sufficient conditions for delay-independent global stability of $(i)$ continuous-time positive systems whose vector fields are cooperative and homogeneous of {\em arbitrary degree} and $(ii)$ discrete-time positive systems whose vector fields are non-decreasing and homogeneous of {\em degree zero}. We demonstrate that such systems are insensitive to a general class of time delays which includes bounded and unbounded time-varying delays.
\item When the asymptotic behaviour of the time delays is known, we obtain conditions to ensure global $\mu$-stability in the sense of \cite{Chen:07,LLiu:08,LLiu:10}. These results allow us to quantify the decay rates of positive systems for various classes of (possibly unbounded) time-varying delays.
\item For bounded delays and a particular class of unbounded delays, we present explicit bounds on the decay rates. These bounds quantify how the magnitude of bounded delays and the rate at which the unbounded delays grow large affect the decay rate.
\item We also show that discrete-time positive systems whose vector fields are non-decreasing and homogeneous of {\em degree greater than zero} are locally asymptotically stable under delay-independent global stability conditions that we have derived.
\end{itemize}

The remainder of the paper is organized as follows. In Section \ref{sec:preliminaries}, we introduce the notation and review some preliminaries that are essential for the development of the results in this paper. Our main results for continuous- and discrete-time nonlinear positive systems are stated in Sections \ref{sec:Continuous-Time Case} and \ref{sec:Discrete-Time Case}, respectively. An illustrative example, justifying the validity of our results, is presented in Section \ref{sec:examples}. Finally, concluding remarks are given in Section \ref{sec:conclusions}.

%
%

\section{Notation and Preliminaries}\label{sec:preliminaries}

\subsection{Notation}

Vectors are written in bold lower case letters and matrices in capital letters. We let $\mathbb{R}$, $\mathbb{N}$, and $\mathbb{N}_0$ denote the set of real numbers, natural numbers, and the set of natural numbers including zero, respectively. The nonnegative orthant of the {\em n}-dimensional real space $\mathbb{R}^n$ is represented by $\mathbb{R}^n_+$. The {\em i}th component of a vector $\x\in \mathbb{R}^n$ is denoted by $x_i$, and the notation $\x\geq \y$ means that $x_i\geq y_i$ for all components $i$. If $\v$ is a vector in $\mathbb{R}^n$, the notation $\v > \bnull$ indicates that all components of $\v$ are positive. Given a vector $\v>\bnull$, the weighted $l_\infty$ norm is defined by
\begin{eqnarray*}
\|\x\|_\infty^{\v}=\max_{1\leq i\leq n}{\frac{|x_i|}{v_i}}.
\end{eqnarray*}
For a matrix $A\in \mathbb{R}^{n\times n}$, $a_{ij}$ denotes the real-valued entry in row $i$ and column $j$. A matrix $A\in \mathbb{R}^{n\times n}$ is said to be {\em nonnegative} if $a_{ij}\geq 0$ for all $i$ and $j$. It is called {\em Metzler} if $a_{ij}\geq 0$ for all $i\neq j$. Given an $n$-tuple $\r = (r_1,\ldots,r_n)$ of positive real numbers and $\lambda> 0$,  the {\em dilation map} $\delta^{\r}_{\lambda}(\x) : \mathbb{R}^n \rightarrow \mathbb{R}^n$ is given by
\begin{eqnarray*}
\delta^{\r}_{\lambda}\bigl(\x\bigr)=\bigl(\lambda^{r_1}x_1,\ldots,\lambda^{r_n}x_n\bigr).
\end{eqnarray*}
If $\r =(1,\ldots,1)$, the dilation map is called the {\em standard dilation map}. For a real interval $[a,b]$, $\mathcal{C}\bigl([a,b],\mathbb{R}^{n}\bigr)$ denotes the space of all real-valued continuous functions on $[a,b]$ taking values in $\mathbb{R}^{n}$. The upper-right Dini-derivative of a continuous function $h:\mathbb{R}\rightarrow \mathbb{R}$ at $t=t_0$ is defined by
\begin{eqnarray*}
D^+h(t)\bigl|_{t=t_0}=\lim_{\Delta\rightarrow 0^+}\sup\frac{h(t_0+\Delta)-h(t_0)}{\Delta},
\end{eqnarray*}
where $\Delta\rightarrow 0^+$ means that $\Delta$ approaches zero from the right-hand side.

\subsection{Preliminaries}

Next, we review the key definitions and results necessary for developing the main results of this paper. We start with the definition of {\em cooperative} vector fields.

\begin{definition}
A continuous vector field $f:\mathbb{R}^{n} \rightarrow \mathbb{R}^{n}$ which is continuously differentiable on $\mathbb{R}^{n}\backslash\{\bnull\}$ is said to be cooperative if the Jacobian matrix ${\partial f}/{\partial x}$ is Metzler for all $\x\in\mathbb{R}^{n}_+\backslash\{\bnull\}$.
\end{definition}
Cooperative vector fields satisfy the following property.

\begin{proposition}\textup{\textbf{\cite[Remark 3.1.1]{Smith:95}}}
\label{Proposition 0}
Let $f:\mathbb{R}^{n} \rightarrow \mathbb{R}^{n}$ be cooperative. For any two vectors $\x$ and $\y$ in $\mathbb{R}^{n}_+\backslash\{\bnull\}$ with $x_i=y_i$ and $\x\geq \y$, we have $f_i(\x)\geq f_i(\y)$.
\end{proposition}
The following definition introduces {\em homogeneous} vector fields.

\begin{definition}
For any $p\geq 0$, the vector field $f : \mathbb{R}^n \rightarrow \mathbb{R}^n$  is said to be homogeneous of degree $p$ with respect to the dilation map $\delta^{\r}_{\lambda}(\x)$  if
\begin{eqnarray*}
\f\bigl(\delta^{\r}_{\lambda}(\x)\bigr)=\lambda^{p}\delta^{\r}_{\lambda}\bigl(\f(\x)\bigr),\quad \forall \x\in\mathbb{R}^n,\; \forall \lambda> 0.
\end{eqnarray*}
\end{definition}
Finally, we define {\em non-decreasing} vector fields.

\begin{definition}
A vector field $g:\mathbb{R}^{n} \rightarrow \mathbb{R}^{n}$ is said to be non-decreasing on $\mathbb{R}^{n}_+$ if $\g(\x)\geq \g(\y)$ for any $\x,\y\in\mathbb{R}^{n}_+$ such that $\x\geq \y$.
\end{definition}

%
%

\section{Continuous-Time Homogeneous Cooperative Systems}\label{sec:Continuous-Time Case}

\subsection{Problem Statement}

Consider the continuous-time dynamical system
\begin{eqnarray}
\label{System 3}
{\mathcal G:}
& \left\{
\begin{array}[l]{ll}
\dot{\x}\bigl(t\bigr)=\f\bigl(\x(t)\bigr)+\g\bigl(\x(t-\tau(t))\bigr),&t\geq 0,\\[0.05cm]
\x\bigl(t\bigr)=\bphi\bigl(t\bigr),&t\in[-\tau_{\max},0],
\end{array}
\right.
\end{eqnarray}
where $\x(t)\in\mathbb{R}^{n}$ is the state variable, and $f,g:\mathbb{R}^{n} \rightarrow \mathbb{R}^{n}$ are continuous vector fields on $\mathbb{R}^{n}$, continuously differentiable on $\mathbb{R}^{n}\backslash\{\bnull\}$, and such that $\f(\bnull)=\g(\bnull)=\bnull$. Here, $\bphi(\cdot)\in \mathcal{C}\bigl([-\tau_{\max},0],\mathbb{R}^{n}\bigr)$ is the vector-valued function specifying the initial condition of the system, and $\tau(\cdot)$ is the time-varying delay which satisfies the following assumption:

\begin{assumption}
\label{Assumption 5}
The delay $\tau:\mathbb{R}_+\rightarrow \mathbb{R}_+$ is continuous with respect to time and satisfies
\begin{equation}
\label{Unbounded Assumption 1}
\lim_{t \rightarrow +\infty} t-\tau(t)=+\infty.
\end{equation}
\end{assumption}

Note that $\tau(t)$ is not necessarily continuously differentiable and no restriction on its derivative (such as $\dot{\tau}(t)<1$) is imposed. Condition (\ref{Unbounded Assumption 1}) implies that as $t$ increases, $\tau(t)$ grows slower than time itself. This constraint on time delays is not restrictive and typically satisfied in real-world applications. For example, the continuous-time power control algorithm for a wireless network consisting of $n$ mobile users can be described by
\begin{equation}
\label{Power Control}
\dot{x_i}\bigl(t\bigr)=-x_i\bigl(t\bigr)+\sum_{\substack{j=1\\j\neq i}}^n a_{ij}x_j\bigl(t-\tau(t)\bigr),\quad i=1,\ldots,n.
\end{equation}
Here, $x_i$ represents the transmit power of user $i$, and $a_{ij}$ are nonnegative constants~\cite{2008:constantdelays,Hamid:13a ,2012:unconditional}. If $\tau(t)$ satisfies (\ref{Unbounded Assumption 1}), then given any time $t_1\geq 0$, there exists a time $t_2>t_1$ such that
\begin{eqnarray*}
t-\tau(t)\geq t_1,\quad \forall t\geq t_2.
\end{eqnarray*}
This simply means that given any time $t_1$, information about which transmit power each user has applied prior to $t_1$ will be received by every other user before a sufficiently long time $t_2$ and not be used in the state evolution of (\ref{Power Control}) after $t_2$. In other words, state information eventually propagates to all other users in the network and old information is eventually purged from the network. In the power control problem,  Assumption \ref{Assumption 5} is always satisfied unless the communication between users is totally lost during a semi-infinite time interval.

Note that all bounded delays, irrespective of whether they are constant or time-varying, satisfy Assumption \ref{Assumption 5}. Moreover, delays satisfying (\ref{Unbounded Assumption 1}) may be unbounded. Consider the following particular class of unbounded delays which was studied in \cite{TChen:07,Liu:2011}.

\begin{assumption}
\label{Assumption 5.1}
There exist $T>0$ and a scalar $0 \leq \alpha<1$ such that
\begin{equation}
\label{Unbounded Assumption 1.1}
\sup_{t>T}\frac{\tau(t)}{t}=\alpha.
\end{equation}
\end{assumption}

One can easily verify that (\ref{Unbounded Assumption 1.1}) implies (\ref{Unbounded Assumption 1}). However, the next example shows that the converse does not, in general, hold. Hence, Assumption \ref{Assumption 5.1} is a special case of Assumption \ref{Assumption 5}.

\begin{example}
Let $\tau(t)=t-\ln(t+1)$ for $t\geq 0$. Since
\begin{eqnarray*}
\lim_{t \rightarrow +\infty} t-\tau(t)&=&\lim_{t \rightarrow +\infty}\ln(t+1)=+\infty,\\
\lim_{k \rightarrow +\infty} \frac{\tau(t)}{t}&=&\lim_{t \rightarrow +\infty} \frac{t-\ln(t+1)}{t}=1,
\end{eqnarray*}
it is clear that (\ref{Unbounded Assumption 1}) holds while (\ref{Unbounded Assumption 1.1}) does not hold.
\end{example}

\begin{remark}
\label{Remark Unbounded Delay}
Assumption \ref{Assumption 5} implies that there exists a sufficiently large $T_0>0$ such that $t-\tau(t)>0$ for all $t>T_0$. Define
\begin{equation*}
\tau_{\max}=-\inf_{0\leq t\leq T_0}\biggl\{t-\tau(t)\biggr\}.
\end{equation*}
Since $\tau_{\max}\geq 0$ is bounded, it follows that for any delay satisfying Assumption \ref{Assumption 5}, even if it is unbounded, the initial condition $\bphi(\cdot)$ is defined on a bounded set $[-\tau_{\max},0]$.
\end{remark}

In this section, we study delay-independent stability of nonlinear systems of the form (\ref{System 3}) which are {\em positive} defined as follows.

\begin{definition}
System $\mathcal{G}$ given by (\ref{System 3}) is said to be positive if for every nonnegative initial condition $\bphi(\cdot)\in \mathcal{C}\bigl([-\tau_{\max},0],\mathbb{R}^{n}_+\bigr)$, the corresponding state trajectory is nonnegative, that is $\x(t)\in \mathbb{R}_+^{n}$ for all $t\geq 0$.
\end{definition}

The following result provides a sufficient condition for positivity of $\mathcal{G}$.

\begin{proposition}
\label{Proposition 2}
Consider the time-delay system $\mathcal{G}$ given by (\ref{System 3}). If the following condition holds:
\begin{align}
\label{Proposition 2.0}
\begin{split}
\forall i\in\{1,\ldots,n\},\;&\forall \x \in \mathbb{R}^n_+\;:\;x_i=0 \Rightarrow f_i(\x)\geq 0,\\
&\forall \x \in \mathbb{R}^n_+, \hspace{1.95cm} \g(\x)\geq 0,
\end{split}
\end{align}
then $\mathcal{G}$ is positive.
\end{proposition}

\begin{proof}
See \cite{Feyzmahdavian:14-2}.
\end{proof}

Note that the nonnegativity of the initial condition is essential for ensuring positivity of the state evolution of the  system $\mathcal{G}$ given by (\ref{System 3}). In other words, when $\bphi(\cdot)\geq \bnull$ is not satisfied, $\x(t)$ may not stay in the positive orthant even if the conditions of Proposition \ref{Proposition 2} hold.

In \cite[Proposition 3.1]{Haddad:10}, it was shown that for nonzero constant delays, the sufficient condition in Proposition \ref{Proposition 2} is also necessary, \textit{i.e.}, a system $\mathcal{G}$ given by (\ref{System 3}) with $\tau(t)=\tau_{\max}>0$, $t\geq 0$, is positive if and only if (\ref{Proposition 2.0}) holds. However, the condition is not necessary when we allow for time-varying delays, as the next example shows.

\begin{example}
Consider a continuous-time linear system described by (\ref{System 3}) with
\begin{equation}
\label{Example 0}
\f(x_1,x_2)=\begin{bmatrix} 1 & 0\\[0.05cm] -1 & 0 \end{bmatrix}\begin{bmatrix} x_1 \\ x_2 \end{bmatrix},\quad \g(x_1,x_2)=\begin{bmatrix}0 & 0\\[0.05cm]  e & 0 \end{bmatrix}\begin{bmatrix} x_1 \\ x_2 \end{bmatrix},
\end{equation}
where $e$ is the base of the natural logarithm, and let the time-varying delay be
\begin{equation}
\label{Example 0.1}
\tau(t)=
\begin{cases}
0, & 0\leq t\leq 1, \\
t-1,  & 1\leq t\leq 2, \\
1, & 2\leq t.
\end{cases}
\end{equation}
Note that $0\leq \tau(t)\leq 1$ for all $t\geq 0$. The solution to (\ref{Example 0}) is given by
\begin{eqnarray*}
x_1(t)&=&x_1(0)e^t,\quad \hspace{3.83cm} 0\leq t,\\
x_2(t)&=&
\begin{cases}
x_2(0)+(e-1)(e^t-1)x_1(0), & 0\leq t\leq 1, \\
x_2(0)+(e^2t-e^t+1-e)x_1(0),  & 1\leq t\leq 2, \\
x_2(0)+(e^2-e+1)x_1(0), & 2\leq t.
\end{cases}
\end{eqnarray*}
It is straightforward to verify that $\x(t)\geq \bnull$ for every nonnegative initial condition $\x(0)=(x_1(0),x_2(0))$, and hence the linear system (\ref{Example 0}) with the bounded time-varying delay (\ref{Example 0.1}) is positive. However, the sufficient condition given in Proposition~\ref{Proposition 2} is not satisfied in this example, since $x_2=0$ does not imply $f_2(\x)\geq 0$, $\forall\x \in \mathbb{R}^2_+$ (take, for example, $f_2(1,0)=-1<0$).
\end{example}

From this point on, vector fields $\f$ and $\g$ satisfy Assumption \ref{Assumption 6}.

\begin{assumption}
\label{Assumption 6}
The following properties hold:
\begin{enumerate}
\item $\f$ is cooperative and $\g$ is non-decreasing on $\mathbb{R}^{n}_+$;
\item $\f$ and $\g$ are homogeneous of degree $p$ with respect to the dilation map $\delta^{\r}_{\lambda}(\x)$.
\end{enumerate}
\end{assumption}

A  system $\mathcal{G}$ given by (\ref{System 3}) satisfying Assumption \ref{Assumption 6} is called {\em homogeneous cooperative}. According to Propositions \ref{Proposition 0} and \ref{Proposition 2}, since $\f(\bnull)=\g(\bnull)=\bnull$, Assumption \ref{Assumption 6}.1) ensures the positivity of homogeneous cooperative systems. The model of some physical systems fall within this class of positive systems. For example, continuous-time linear and several nonlinear power control algorithms for wireless networks are described by homogeneous cooperative systems \cite{Boche:08,Feyzmahdavian:12,Feyzmahdavian:14-3}.

While the stability of general dynamical systems may depend on the magnitude and variation of the time delays, homogeneous cooperative systems have been shown to be insensitive to constant delays~\cite{Vahid:10}. More precisely, the homogeneous cooperative system (\ref{System 3}) with a constant delay $\tau(t)=\tau_{\max}$, $t\geq 0$, is globally asymptotically stable for all $\tau_{\max}\geq 0$ if and only if the undelayed system $(\tau_{\max}= 0)$ is globally asymptotically stable. The main goal of this section is to $(i)$ determine whether a similar delay-independent stability result holds for homogeneous cooperative systems with  time-varying delays satisfying Assumption \ref{Assumption 5}; and to $(ii)$ give explicit estimates of the decay rate for different classes of time delays (e.g., bounded delays, unbounded delays satisfying Assumption \ref{Assumption 5.1}, etc.).

\subsection{Asymptotic Stability of Homogeneous Cooperative Systems}

\begin{figure}[h]
\centering
\includegraphics [width=0.85\columnwidth]{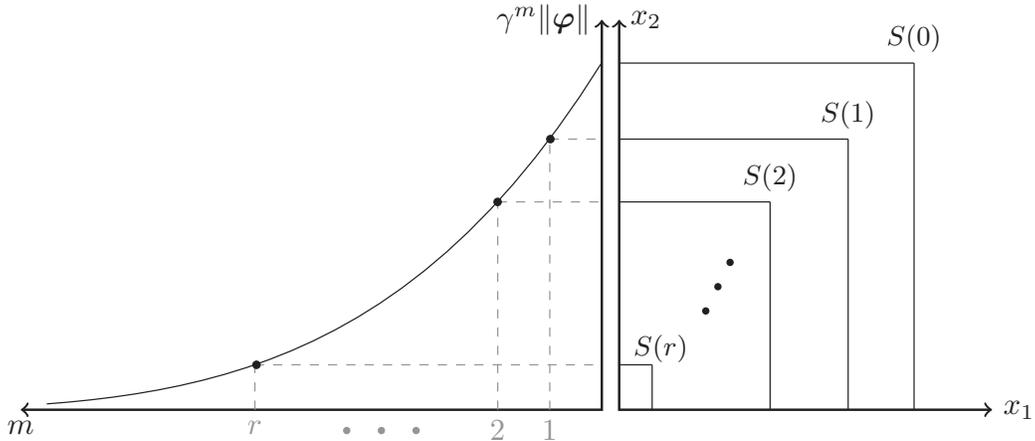}
\caption{Level curves of the Lyapunov function $V(\x)$ in the two-dimensional case. The key idea behind the proof of Theorem~\ref{Theorem 3} is that $\bphi(\cdot)$ is initially within the set $S(0)$ and at some time $t_1 \geq 0$ eventually $\x(t)$ enters and stays within the set $S(1)$ for all $t\geq t_1$; moreover, as $t$ increases further, $\x(t)$ sequentially moves into other sets.}
\label{Figure 1}
\end{figure}

The following theorem establishes a necessary and sufficient condition for delay-independent asymptotic stability of homogeneous cooperative systems with time-varying delays satisfying Assumption \ref{Assumption 5}. Our proof (which is conceptually related to the Lyapunov stability theorem) uses the Lyapunov function
\begin{equation*}
V(\x)=\max_{1\leq i\leq n}\left({\frac{x_i}{v_i}}\right)^{\frac{r_{\max}}{r_i}},
\end{equation*}
where $\v>\bnull$, and $r_{\max}=\max_{1\leq i\leq n}r_i$, to define sets
\begin{equation}
\label{Level Sets}
S(m)=\biggl\{\x\in \mathbb{R}^n_+  \;\bigl|\; V(\x) \leq  \gamma^m\|\bphi\| \biggr\},\quad m\in \mathbb{N}_0,
\end{equation}
where $0\leq \gamma<1$, and
\begin{equation}
\label{Initial Condition}
\|\bphi\|=\sup_{-\tau_{\max} \leq s\leq 0}V(\bphi(s)),
\end{equation}
and shows that for each $m$, there exists $t_m\geq 0$ such that $\x(t)\in S(m)$ for all $t\geq t_m$. In other words, the system state will enter each set $S(m)$ at some time $t_m$ and remain in the set for all future times. Since the sets are nested, \textit{i.e.},
\begin{equation*}
S(0)\supset \cdots \supset S(m)\supset S(m+1)\supset \cdots,
\end{equation*}
the state will move sequentially from set $S(m)$ to $S(m+1)$, cf. Figure \ref{Figure 1}.
Thus, the sets play a similar role as level sets of the Lyapunov function $V(\x)$. Note that when $\f$ and $\g$ are homogeneous with respect to the standard dilation map, $V(\x)=\|\x \|_{\infty}^{\v}$, which is often used in analysis of positive linear systems~\cite{Rantzer:11}.

\begin{theorem}
\label{Theorem 3}
For the homogeneous cooperative system $\mathcal{G}$ given by (\ref{System 3}), suppose that Assumption \ref{Assumption 5} holds. Then, the following statements are equivalent.
\begin{itemize}
\item[$(i)$] There exists a vector $\v>\bnull$ such that
\begin{equation}
\label{Theorem 3-0}
\f(\v)+\g(\v)<\bnull.
\end{equation}
\item[$(ii)$] Homogeneous cooperative system $\mathcal{G}$ is globally asymptotically stable for every nonnegative initial condition $\bphi(\cdot)\in\mathcal{C}([-\tau_{\max},0],\mathbb{R}^{n}_+)$, and for all time delays satisfying Assumption \ref{Assumption 5}.
\item[$(iii)$] Homogeneous cooperative system $\mathcal{G}$ without delay $(\tau(t)=0,\;t\geq 0)$ is globally asymptotically stable for all nonnegative initial conditions.
\end{itemize}
\end{theorem}

\begin{proof}
See \cite{Feyzmahdavian:14-2}.
\end{proof}

The stability condition (\ref{Theorem 3-0}) does not include any information on the magnitude and variation of delays, so it ensures {\em delay-independent} stability.  According to Theorem \ref{Theorem 3}, the homogeneous cooperative system~$\mathcal{G}$ given by (\ref{System 3}) is globally asymptotically stable for all time delays satisfying Assumption \ref{Assumption 5} if and only if the corresponding delay-free system is globally asymptotically stable. This property is very useful in practical applications, since the delays may not be easy to model in detail.

Note that Theorem \ref{Theorem 3} can be easily extended to homogeneous cooperative systems with multiple delays of the form
\begin{equation*}
\dot{\x}\bigl(t\bigr)=\f\bigl(\x(t)\bigr)+\sum _{q=1}^s\g_q\bigl(\x(t-\tau_q(t))\bigr).
\end{equation*}
Here, $s\in\mathbb{N}$, $\f$ is cooperative and homogeneous, $\g_q$ for $q=1,\ldots,s$ are homogenous and non-decreasing on $\mathbb{R}^{n}_+$, and $\tau_q(t)$ satisfy Assumption \ref{Assumption 5}. In this case, the stability condition (\ref{Theorem 3-0}) becomes
\begin{equation*}
\f(\v)+\sum _{q=1}^s \g_q(\v)<\bnull.
\end{equation*}

\begin{remark}
In \cite{Feyzmahdavian:14}, it has been shown that if $\f$ and $\g$ are homogeneous of degree zero with respect to the standard dilation map, the homogeneous cooperative system (\ref{System 3}) is insensitive to bounded time-varying delays. In this work, we extend this result to cooperative systems that are homogeneous of any degree with respect to an arbitrary dilation map. Moreover, we impose minimal restrictions on time delays and establish insensitivity of homogeneous cooperative systems to the general class of delays described by Assumption \ref{Assumption 5}, which includes bounded delays as a special case.
\end{remark}

\subsection{Decay Rates of Homogeneous Cooperative Systems}

Theorem \ref{Theorem 3} is concerned with the {\em asymptotic} stability of homogeneous cooperative systems with time-varying delays. However, there are processes and applications for which it is desirable that the system has a certain decay rate. Loosely speaking, the system has to converge quickly enough to the equilibrium. Hence, it is important to investigate the impact of delays on the decay rate of such systems. In this section, we characterize how time delays affect the decay rate of the homogeneous cooperative system $\mathcal{G}$ given by (\ref{System 3}). Before stating the main result, we provide the definition of $\mu$-stability, introduced in \cite{Chen:07}, for continuous-time systems.

\begin{definition}
Suppose that $\mu:\mathbb{R}_+\rightarrow \mathbb{R}_+$ is a non-decreasing function satisfying $\mu(t)\rightarrow+\infty$ as $t\rightarrow+\infty$. System $\mathcal{G}$ given by (\ref{System 3}) is said to be globally $\mu$-stable if there exists a constant $M>0$ such that for any initial function $\bphi(\cdot)$, the solution $\x(t)$ satisfies
\begin{equation*}
\|\x(t)\|\leq \frac{M}{\mu(t)},\quad t>0,
\end{equation*}
where $\|\cdot\|$ is some norm on $\mathbb{R}^n$.
\end{definition}

This definition can be regarded as a unification of several types of stability. For example, if~$\mu(t)=e^{\eta t}$ with $\eta>0$, the $\mu$-stability becomes {\em exponential stability}; and when $\mu(t)=t^\xi$ with $\xi>0$, then the $\mu$-stability becomes {\em power-rate stability}~\cite{Chen:07,LLiu:08,LLiu:10}.

Global $\mu$-stability of homogenous cooperative systems can be verified using the following theorem.

\begin{theorem}
\label{Theorem 3-1}
Consider the homogeneous cooperative system $\mathcal{G}$ given by (\ref{System 3}). Suppose that Assumption~\ref{Assumption 5} holds, and that there is a vector $\v>\bnull$ satisfying
\begin{equation}
\label{Theorem 3-1-0}
\f(\v)+\g(\v)<\bnull.
\end{equation}
If there exists a function $\mu:\mathbb{R}_+\rightarrow \mathbb{R}_+$ such that the following conditions hold:
\begin{itemize}
\item[$(i)$]  $\mu(t)>0$, for all $t>0$;
\item[$(ii)$]  $\mu(t)$ is a non-decreasing function;
\item[$(iii)$]  $\lim_{t\rightarrow +\infty} \mu(t)=+\infty$;
\item[$(iv)$]  For each $i\in\{1,\ldots,n\}$,
\begin{equation*}
\left(\frac{r_{\max}}{r_i}\right)\left(\left(\frac{f_i(\v)}{v_i}\right)+
\left(\lim_{t\rightarrow \infty}\frac{\mu(t)}{\mu(t-\tau(t))}\right)^{\frac{r_i+p}{r_{\max}}}\left(\frac{g_i(\v)}{v_i}\right)\right)+\lim_{t\rightarrow \infty}\frac{\dot{\mu}(t)}{\left(\mu(t)\right)^{1-\frac{p}{r_{\max}}}}<0,
\end{equation*}
\end{itemize}
then every solution of $\mathcal{G}$ starting in the positive orthant satisfies
\begin{equation*}
\left({\frac{x_i(t)}{v_i}}\right)^{\frac{r_{\max}}{r_i}}=O\left(\mu^{-1}(t)\right),\quad t\geq 0,
\end{equation*}
for $i=1,\ldots,n$.
\end{theorem}

\begin{proof}
See \cite{Feyzmahdavian:14-2}.
\end{proof}

According to Theorem \ref{Theorem 3-1}, any function $\mu(t)$ satisfying conditions $(i)$--$(iv)$ can be used to estimate the decay rate of homogeneous cooperative systems with time-varying delays satisfying Assumption \ref{Assumption 5}. From condition $(iv)$, it is clear that the asymptotic behaviour of the delay $\tau(t)$ influences the admissible choices for $\mu(t)$ and, hence, the decay bounds that we are able to guarantee. To clarify this statement, we will analyze a few special cases in detail. First, assume that $\tau(t)$ is bounded, \textit{i.e.},
\begin{equation}
\label{Bounded Delay}
0\leq \tau(t)\leq \tau_{\max}, \quad t\geq 0.
\end{equation}
The following result shows that the decay rate of homogeneous cooperative systems of degree $p$ with bounded time-varying delays is upper bounded by an exponential function of time when $p=0$ and by a polynomial function of time when $p>0$.

\begin{corollary}
\label{Corollary 3}
For the homogeneous cooperative system $\mathcal{G}$ given by (\ref{System 3}), suppose that (\ref{Bounded Delay}) holds and that there exists a vector $\v>\bnull$ satisfying (\ref{Theorem 3-1-0}).
\begin{itemize}
\item[(i)] If $\f$ and $\g$ are homogeneous of degree $p=0$, then $\mathcal{G}$ is globally exponentially stable. In particular,
\begin{equation}
\label{Corollary 3-11}
\left({\frac{x_i(t)}{v_i}}\right)^{\frac{r_{\max}}{r_i}}= O\left(e^{-\eta t}\right),\quad t\geq 0,
\end{equation}
where $0<\eta<\min_{1\leq i\leq n}\eta_i$, and $\eta_i$ is the positive solution of the equation
\begin{equation}
\label{Corollary 3-1}
\left(\frac{r_{\max}}{r_i}\right)\left(\left(\frac{f_i(\v)}{v_i}\right)+\biggl(e^{\eta_i\tau_{\max}}\biggr)^{\frac{r_i}{r_{\max}}}
\left(\frac{g_i(\v)}{v_i}\right)\right)+\eta_i=0.
\end{equation}
\item[(ii)] If $\f$ and $\g$ are homogeneous of degree $p>0$, the solution $\x(t)$ of $\mathcal{G}$ satisfies
\begin{equation}
\label{Corollary 3-1-1-0}
\left({\frac{x_i(t)}{v_i}}\right)^{\frac{r_{\max}}{r_i}}=O\left((\theta t+1)^{\frac{-r_{\max}}{p}}\right),\quad t\geq 0,
\end{equation}
where $0<\theta<\min\{\frac{1}{\tau_{\max}},\min_{1\leq i\leq n}\theta_i\}$, and $\theta_i$ is the positive solution to
\begin{equation}
\label{Corollary 3-1-1}
\frac{f_i(\v)}{v_i}+\frac{g_i(\v)}{v_i}+\theta_i\frac{r_{i}}{p}=0.
\end{equation}
\end{itemize}
\end{corollary}

\begin{proof}
See \cite{Feyzmahdavian:14-2}.
\end{proof}

\begin{remark}
\label{Remark:1-3}
Equation (\ref{Corollary 3-1}) has three parameters: the maximum delay bound $\tau_{\max}$, the positive vector~$\v$ and $\eta_i$.
For any fixed $\tau_{\max}\geq 0$ and any fixed $\v>\bnull$ satisfying (\ref{Theorem 3-1-0}), the left-hand side of (\ref{Corollary 3-1}) is smaller than the right-hand side for $\eta_i=0$, and strictly monotonically increasing in $\eta_i>0$. Therefore, (\ref{Corollary 3-1}) has always a unique positive solution $\eta_i$. By a similar argument, equation (\ref{Corollary 3-1-1}) also admits a unique positive solution $\theta_i$.
\end{remark}

While the stability of homogeneous cooperative systems with delays satisfying Assumption \ref{Assumption 5} may, in general, only be asymptotic, Corollary \ref{Corollary 3} demonstrates that if the delays are bounded, we can guarantee certain decay rates. We will now establish similar decay bounds for unbounded delays satisfying Assumption \ref{Assumption 5.1}.

\begin{corollary}
\label{Corollary 4}
Consider the homogeneous cooperative system $\mathcal{G}$ given by (\ref{System 3}). Suppose that Assumption~\ref{Assumption 5.1} holds and that there exists a vector $\v>\bnull$ satisfying (\ref{Theorem 3-1-0}). Then, $\mathcal{G}$ is globally power-rate stable. Moreover,
\begin{itemize}
\item[(i)] if $\f$ and $\g$ are homogeneous of degree $p=0$, the solution $\x(t)$ of $\mathcal{G}$ satisfies
\begin{equation*}
\left({\frac{x_i(t)}{v_i}}\right)^{\frac{r_{\max}}{r_i}}= O\left(t^{-\xi}\right),\quad t\geq 0,
\end{equation*}
where $0<\xi<\min_{1\leq i\leq n}\xi_i$, and $\xi_i$ is the unique positive solution to
\begin{equation}
\label{Corollary 4-1}
\left(\frac{f_i(\v)}{v_i}\right)+\biggl(\frac{1}{1-\alpha}\biggr)^{\frac{r_i}{r_{\max}}\xi_i}\left(\frac{g_i(\v)}{v_i}\right)=0;
\end{equation}
\item[(ii)] if $\f$ and $\g$ are homogeneous of degree $p>0$, then
\begin{equation*}
\left({\frac{x_i(t)}{v_i}}\right)^{\frac{r_{\max}}{r_i}}=O\left(t^{\frac{-r_{\max}}{p}\beta}\right),\quad t\geq 0,
\end{equation*}
where $\beta\in(0,1)$ is such that
\begin{equation}
\label{Corollary 4-1-1}
\left(\frac{f_i(\v)}{v_i}\right)+\biggl(\frac{1}{1-\alpha}\biggr)^{(1+\frac{r_i}{p})\beta}\left(\frac{g_i(\v)}{v_i}\right)<0,
\end{equation}
holds for all $i$.
\end{itemize}
\end{corollary}

\begin{proof}
See \cite{Feyzmahdavian:14-2}.
\end{proof}

\begin{figure}[h]
\centering
\includegraphics [width=0.4\columnwidth]{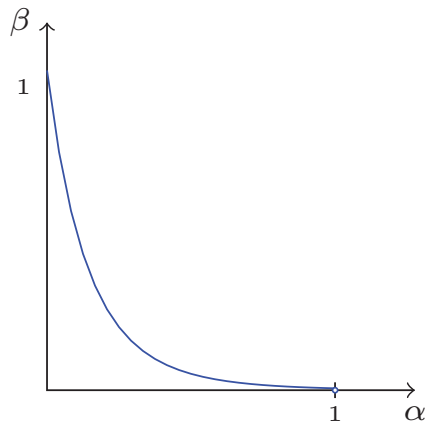}
\caption{Plot of $\beta$ for different values of the parameter $\alpha\in [0,1)$. Clearly, $\beta$ is monotonically decreasing with $\alpha$ and approaches zero as $\alpha$ tends to one.}
\label{Figure 2}
\end{figure}

Corollary \ref{Corollary 4} shows that the decay rate of homogeneous cooperative systems of degree zero with unbounded delays satisfying Assumption \ref{Assumption 5.1} is of order $O(t^{-\xi})$. Equation (\ref{Corollary 4-1}) quantifies how the magnitude of the upper delay bound, $\alpha$, affects $\xi$. Specifically, $\xi_i$ is monotonically decreasing with $\alpha$ and approaches zero as $\alpha$ tends to one. By similar reasoning, $\beta$, on which the guaranteed decay rate of homogeneous cooperative systems of degree greater than zero depends, in equation (\ref{Corollary 4-1-1}) approaches zero as $\alpha$ tends to one (see Figure \ref{Figure 2}).
Hence, while the homogeneous cooperative system (\ref{System 3}) remains power-rate stable for arbitrary unbounded delays satisfying Assumption \ref{Assumption 5.1}, the decay rate deteriorates with increasing $\alpha$. This means that the guaranteed convergence rates get increasingly slower as delays are allowed to grow quicker when $t\rightarrow \infty$.

\subsection{A Special Case: Continuous-Time Positive Linear Systems}

Let $\f(\x)=A\x$ and $\g(\x)=B\x$ such that $A\in\mathbb{R}^{n\times n}$ is Metzler and $B\in\mathbb{R}^{n\times n}$ is nonnegative. Then, the homogeneous cooperative system (\ref{System 3}) reduces to the positive linear system
\begin{eqnarray}
\label{System 4}
{\mathcal G}_{L}:
& \left\{
\begin{array}[l]{ll}
\dot{\x}\bigl(t\bigr)=A\x\bigl(t\bigr)+B\x\bigl(t-\tau(t)\bigr),& t\geq 0,\\[0.1cm]
\x\bigl(t\bigr)=\bphi\bigl(t\bigr), &t\in[-\tau_{\max},0].
\end{array}
\right.
\end{eqnarray}
We then have the following special case of Theorem \ref{Theorem 3}.

\begin{corollary}
\label{Corollary 6}
Consider the positive linear system $\mathcal{G}_L$ given by (\ref{System 4}) where $A$ is Metzler and $B$ is nonnegative. Then, $\mathcal{G}_L$ is globally asymptotically stable for all time delays satisfying Assumption \ref{Assumption 5} if and only if there exists a vector $\v > \bnull$ such that
\begin{equation}
\label{LP}
\bigl(A+B\bigr)\v<\bnull.
\end{equation}
\end{corollary}

Corollary \ref{Corollary 6} shows that if the positive linear system (\ref{System 4}) without delay is stable, it remains asymptotically stable under all bounded and unbounded time-varying delays satisfying Assumption \ref{Assumption 5}. Note that the stability condition (\ref{LP}) is a linear programming (LP) feasibility problem in $\v$ which can be verified numerically in polynomial time.

\begin{remark}
Since $A$ is Metzler and $B$ is nonnegative, $A+B$ is Metzler. It follows from \cite[Proposition 2]{Rantzer:11} that the linear inequality (\ref{LP}) holds if and only if $A+B$ is Hurwitz, \textit{i.e.}, all its eigenvalues have negative real parts.
\end{remark}

While the asymptotic stability of the positive linear system $\mathcal{G}_L$ given by (\ref{System 4}) with time-varying delays satisfying Assumption \ref{Assumption 5} has been investigated in \cite{Sun:12}, the impact of time delays on the decay rate has been missing. Theorem \ref{Theorem 3-1} helps us to find guaranteed decay rates of $\mathcal{G}_L$ for different classes of time delays. Specifically, Corollaries \ref{Corollary 3} and \ref{Corollary 4} show that $\mathcal{G}_L$ is exponentially stable if time-varying delays are bounded, and power-rate stable if delays are unbounded and satisfy Assumption \ref{Assumption 5.1}. Therefore, not only do we extend the result of \cite{Sun:12} to general homogeneous cooperative systems (not necessarily linear), but we also provide explicit bounds on the decay rate of positive linear systems.

\begin{remark}
In \cite[Example 4.5]{Liu:13}, it was shown that a positive linear system with unbounded delays satisfying Assumption \ref{Assumption 5.1} may converge slower than any exponential function. However, an upper bound for the decay rate was not derived in~\cite{Liu:13}. Corollary \ref{Corollary 4} reveals that under Assumption \ref{Assumption 5.1} on delays, the decay rate of positive linear systems is upper bounded by a polynomial function of time.
\end{remark}

%
%

\section{Discrete-time Homogeneous Non-Decreasing Systems}\label{sec:Discrete-Time Case}

\subsection{Problem Statement}

Next, we consider the discrete-time analog of (\ref{System 3}):
\begin{eqnarray}
\label{System 1}
{\Sigma}:
\left\{
\begin{array}[l]{ll}
\x\bigl(k+1\bigr)=\f\bigl(\x(k)\bigr)+\g\bigl(\x(k-d(k))\bigr), &k \in \mathbb{N}_0,\\[0.05cm]
\hspace{0.3cm}\x\bigl(k\bigr)\hspace{0.3cm}=\bphi\bigl(k\bigr), &k\in\{-d_{\max},\ldots,0\}.
\end{array}
\right.
\end{eqnarray}
Here, $\x(k)\in\mathbb{R}^{n}$ is the state variable, $f,g:\mathbb{R}^{n} \rightarrow \mathbb{R}^{n}$ are continuous vector fields with $\f(\bnull)=\g(\bnull)=\bnull$, $d_{\max}\in \mathbb{N}_0$, $\bphi:\{-d_{\max},\ldots,0\} \rightarrow \mathbb{R}^{n}$ is the vector sequence specifying the initial state of the system, and $d(k)$ represents the time-varying delay which satisfies the following assumption.

\begin{assumption}
\label{Assumption 1}
The delay $d:\mathbb{N}_0\rightarrow \mathbb{N}_0$ satisfies
\begin{equation}
\label{Unbounded Assumption:1}
\lim_{k \rightarrow +\infty} k-d(k)=+\infty.
\end{equation}
\end{assumption}

Intuitively, if Assumption \ref{Assumption 1} does not hold, computation of $\x(k)$, even for large values of $k$, may involve the initial condition $\bphi(\cdot)$ and those states near it, and hence $\x(k)$ may not converge to zero as $k\rightarrow \infty$. To avoid this situation, Assumption \ref{Assumption 1} guarantees that old  state information is eventually not used in evaluating (\ref{System 1}).

\begin{remark}
\label{Remark Delay}
Assumption \ref{Assumption 1} implies that there exists a sufficiently large $T_0\in\mathbb{N}_0$ such that $k-d(k)>0$ for all $k>T_0$. Let
\begin{equation*}
d_{\max}=-\inf_{0\leq k\leq T_0}\biggl\{k-d(k)\biggr\}.
\end{equation*}
Clearly, $d_{\max}\in\mathbb{N}_0$ is bounded. It follows that, even for unbounded delays satisfying Assumption \ref{Assumption 1}, the initial condition $\bphi(\cdot)$ is defined on a finite set $\{-d_{\max},\ldots,0\}$.
\end{remark}

\begin{definition}
The system $\Sigma$ given by (\ref{System 1}) is said to be positive if for every nonnegative initial condition $\bphi(\cdot)\in\mathbb{R}^{n}_+$, the corresponding solution is nonnegative, that is $\x(k)\geq \bnull$ for all $k \in \mathbb{N}$.
\end{definition}

Positivity of $\Sigma$ is readily verified using the following result.

\begin{proposition}
\label{Proposition 1}
Consider the time-delay system $\Sigma$ given by (\ref{System 1}). If $\f(\x)\geq \bnull$ and $\g(\x)\geq \bnull$ for all $\x\in\mathbb{R}^{n}_+$, then $\Sigma$ is positive.
\end{proposition}

For nonzero constant delays ($d(k)=d_{\max}>0$, $k\in\mathbb{N}_0$), the sufficient condition in Proposition \ref{Proposition 1} is also necessary \cite[Proposition 3.4]{Haddad:10}. However, the following example shows that this result may not true when delays are time-varying.

\begin{example}
Consider a discrete-time linear system described by (\ref{System 1}) with
\begin{equation*}
f(x)=2x,\;g(x)=-x,\;d(k)=\frac{1}{2}\left(1-(-1)^k\right),\quad k\in\mathbb{N}_0.
\end{equation*}
Since $g(x)<0$ for $x>0$, the sufficient condition given in Proposition \ref{Proposition 1} is not satisfied. However, it is easy to verify that the solution of this system is $x(k)=x(0)$, $k\in\mathbb{N}_0$, and hence $x(k)\geq 0$ for all $x(0)\geq 0$.
\end{example}

In this section, vector fields $\f$ and $\g$ satisfy the next assumption.

\begin{assumption}
\label{Assumption 3}
The following properties hold.
\begin{enumerate}
\item $\f$ and $\g$ are non-decreasing on $\mathbb{R}^{n}_+$;
\item $\f$ and $\g$ are homogeneous of degree $p$ with respect to the dilation map $\delta^{\r}_{\lambda}(\x)$.
\end{enumerate}
\end{assumption}

A  system $\Sigma$ given by (\ref{System 1}) satisfying Assumption \ref{Assumption 3} is called {\em homogeneous non-decreasing}. Since $\f(\bnull)=\g(\bnull)=\bnull$, Assumption \ref{Assumption 3}.1) implies that $\f$ and $\g$ satisfy the conditions of Proposition \ref{Proposition 1}.  Hence, homogeneous non-decreasing systems are positive.

Our main objective in this section is to study delay-independent stability of homogeneous non-decreasing systems of the form (\ref{System 1}) with time-varying delays satisfying Assumption \ref{Assumption 1}.

\subsection{Asymptotic Stability of Homogeneous Non-Decreasing Systems}

The next theorem shows that global asymptotic stability of non-decreasing systems of the form (\ref{System 1}) that are homogeneous of degree zero is insensitive to bounded and unbounded time-varying delays satisfying Assumption \ref{Assumption 1}.

\begin{theorem}
\label{Theorem 1}
For the homogeneous non-decreasing system $\Sigma$ given by (\ref{System 1}), suppose that $\f$ and $\g$ are homogeneous of degree $p=0$. Then, the following statements are equivalent.
\begin{itemize}
\item[$(i)$] There exists a vector $\v>\bnull$ such that
\begin{equation}
\label{Theorem 1-0}
\f(\v)+\g(\v)<\v.
\end{equation}
\item[$(ii)$] $\Sigma$ is globally asymptotically stable for any nonnegative initial conditions and for all bounded and unbounded time-varying delays satisfying Assumption \ref{Assumption 1}.
\item[$(iii)$] $\Sigma$ without delay $(d(k)=0,\;k\in\mathbb{N}_0)$ is globally asymptotically stable for any nonnegative initial conditions.
\end{itemize}
\end{theorem}

\begin{proof}
See \cite{Feyzmahdavian:14-2}.
\end{proof}

Theorem \ref{Theorem 1} provides a test for global asymptotic stability of homogeneous non-decreasing systems of degree zero; if we can demonstrate the existence of a vector $\v>\bnull$ satisfying (\ref{Theorem 1-0}), then the origin is globally asymptotically stable for  all delays satisfying Assumption \ref{Assumption 1}. However, the following example illustrates that (\ref{Theorem 1-0}) is, in general, not a sufficient condition for global asymptotic stability of homogeneous non-decreasing systems of degree greater than zero.

\begin{example}
Consider a discrete-time system described by (\ref{System 1}) with $f(x)=x^2$ and $g(x)=0$. Clearly, $f$ is non-decreasing on $\mathbb{R}_+$ and homogeneous of degree one with respect to the standard dilation map. Since $f(0.5)=0.25<0.5$, (\ref{Theorem 1-0}) holds. However, it is easy to verify that solutions of this system starting from initial conditions $x(0)\geq 1$ do not converge to the origin, \textit{i.e.}, the origin is not globally asymptotically stable.
\end{example}

We now show that under stability condition (\ref{Theorem 1-0}), homogeneous non-decreasing systems of degree greater than zero with time-varying delays satisfying Assumption~\ref{Assumption 1} have a locally asymptotically stable equilibrium point at the origin, \textit{i.e.}, $\x(k)$ converges to the origin as $k\rightarrow \infty$ for sufficiently small initial conditions.

\begin{corollary}
\label{Corollary 5}
For the homogeneous non-decreasing system $\Sigma$ given by (\ref{System 1}) with degree $p>0$, suppose that Assumption \ref{Assumption 1} holds. If there exists a vector $\v>\bnull$ such that (\ref{Theorem 1-0}) holds, then the origin is asymptotically stable with respect to initial conditions satisfying
\begin{equation*}
\bnull \leq \bphi(k)\leq \v,\quad \forall k\in\{-d_{\max},\ldots,0\}.
\end{equation*}
\end{corollary}

\begin{proof}
See \cite{Feyzmahdavian:14-2}.
\end{proof}

\subsection{Decay Rates of Homogeneous Non-Decreasing Systems of degree zero}

The next definition introduces $\mu$-stability for discrete-time systems.

\begin{definition}
Suppose that $\mu:\mathbb{N}\rightarrow \mathbb{R}_+$ is a non-decreasing function satisfying $\mu(k)\rightarrow+\infty$ as $k\rightarrow+\infty$. The system $\Sigma$ given by (\ref{System 1}) is said to be globally $\mu$-stable, if there exists a constant $M>0$ such that for any initial function $\bphi(\cdot)$, the solution $\x(k)$ satisfies
\begin{equation*}
\|\x(k)\|\leq \frac{M}{\mu(k)},\quad k\in\mathbb{N},
\end{equation*}
where $\|\cdot\|$ is some norm on $\mathbb{R}^n$.
\end{definition}

Paralleling our continuous-time results, global $\mu$-stability of homogeneous non-decreasing systems of degree zero with time-varying delays can be established using the following theorem.

\begin{theorem}
\label{Theorem 1-2}
Consider the homogeneous non-decreasing system $\Sigma$ given by~(\ref{System 1}) with degree $p=0$. Suppose that Assumption \ref{Assumption 1} holds, and that there is a vector $\v>\bnull$ satisfying
\begin{equation}
\label{Theorem 1-2-0}
\f(\v)+\g(\v)<\v.
\end{equation}
If there exists a function $\mu:\mathbb{N}\rightarrow \mathbb{R}_+$ such that the following conditions hold:
\begin{itemize}
\item[$(i)$] $\mu(k)>0$, for all $k\in\mathbb{N}$;
\item[$(ii)$] $\mu(k+1)\geq\mu(k)$, for all $k\in\mathbb{N}$;
\item[$(iii)$] $\lim_{k\rightarrow +\infty} \mu(k)=+\infty$;
\item[$(iv)$] For each $i\in\{1,\ldots,n\}$,
\begin{equation*}
\left(\lim_{k\rightarrow \infty}\frac{\mu(k+1)}{\mu(k)}\right)^{\frac{r_i}{r_{\max}}}\left(\frac{f_i(\v)}{v_i}\right)+
\left(\lim_{k\rightarrow \infty}\frac{\mu(k+1)}{\mu(k-d(k))}\right)^{\frac{r_i}{r_{\max}}}\left(\frac{g_i(\v)}{v_i}\right)<1,
\end{equation*}
\end{itemize}
then every solution of $\Sigma$ starting in the positive orthant satisfies
\begin{equation*}
\left({\frac{x_i(k)}{v_i}}\right)^{\frac{r_{\max}}{r_i}}=O\left(\mu^{-1}(k)\right),\quad k\in \mathbb{N},
\end{equation*}
for $i=1,\ldots,n$.
\end{theorem}

\begin{proof}
See \cite{Feyzmahdavian:14-2}.
\end{proof}

Theorem \ref{Theorem 1-2} allows us to establish convergence rates of homogeneous non-decreasing systems of degree zero under various classes of time-varying delays. We give the following result.

\begin{corollary}
\label{Corollary 1}
For the homogeneous non-decreasing system $\Sigma$ given by (\ref{System 1}) with degree $p=0$, suppose that there exists a vector $\v>\bnull$ satisfying (\ref{Theorem 1-2-0}), and that there exist $T\in\mathbb{N}$ and a scalar $0 \leq \alpha<1$ such that
\begin{equation}
\label{Unbounded Assumption:2}
\sup_{k>T}\frac{d(k)}{k}=\alpha.
\end{equation}
Let $\xi_i$ be the unique positive solution of the equation
\begin{equation}
\label{Theorem 1-3-0}
\left(\frac{f_i(\v)}{v_i}\right)+\left(\frac{1}{1-\alpha}\right)^{\frac{r_i}{r_{\max}}\xi_i}\left(\frac{g_i(\v)}{v_i}\right)=1,\quad i=1,\ldots,n.
\end{equation}
Then, $\Sigma$ is globally power-rate stable for any nonnegative initial conditions and for any unbounded delays satisfying (\ref{Unbounded Assumption:2}). In particular,
\begin{equation*}
\left({\frac{x_i(k)}{v_i}}\right)^{\frac{r_{\max}}{r_i}}=O\left(k^{-\xi}\right),\quad k\in \mathbb{N},
\end{equation*}
where $0<\xi<\min_{1\leq i\leq n}\xi_i$.
\end{corollary}

\subsection{A Special Case: Discrete-Time Positive Linear Systems}

We now discuss delay-independent stability of a special case of (\ref{System 1}), namely discrete-time positive linear systems of the form
\begin{eqnarray}
\label{System 2}
{\Sigma}_L:
& \left\{
\begin{array}[l]{ll}
{\x}\bigl(k+1\bigr)=A\x\bigl(k\bigr)+B\x\bigl(k-d(k)\bigr), &k\in \mathbb{N}_0,\\[0.05cm]
\hspace{0.3cm}\x\bigl(k\bigr)\hspace{0.3cm}=\bphi\bigl(k\bigr),&k\in\{-d_{\max},\ldots,0\}.
\end{array}
\right.
\end{eqnarray}
In terms of (\ref{System 1}), $\f(\x)=A\x$ and $\g(\x)=B\x$. It is easy to verify that if $A,B\in\mathbb{R}^{n\times n}$ are nonnegative matrices, Assumption \ref{Assumption 3} is satisfied. Therefore, Theorem \ref{Theorem 1} can help us to derive a necessary and sufficient condition for delay-independent stability of (\ref{System 2}). Specifically, we note the following.

\begin{corollary}
\label{Theorem 2}
Consider the discrete-time positive linear system $\Sigma_L$ given by~(\ref{System 2}) where $A$ and $B$ are nonnegative. Then, there exists a vector $\v > \bnull$ such that
\begin{equation}
\label{LP-d}
\bigl(A+B\bigr)\v<\v,
\end{equation}
if and only if $\Sigma_L$ is globally asymptotically stable for all time delays satisfying Assumption \ref{Assumption 1}.
\end{corollary}

\begin{remark}
For the positive linear system (\ref{System 2}), $A$ and $B$ are nonnegative, so $A+B$ is also nonnegative. According to property of nonnegative matrices \cite{Berman:79},\cite[Proposition 1]{Rantzer:11}, there exists a vector $\v>\bnull$ satisfying (\ref{LP-d}) if and only if all eigenvalues of $A+B$ are strictly inside the unit circle.
\end{remark}

\begin{remark}
In \cite{Feyzmahdavian:13}, it was shown that discrete-time positive linear systems are insensitive to time delays satisfying Assumption \ref{Assumption 1}. Theorem \ref{Theorem 1} shows that a similar delay-independent stability result holds for nonlinear positive systems whose vector fields are non-decreasing and homogeneous of degree zero. Furthermore, the impact of various classes of time delays on the convergence rate of positive linear systems has been missing in \cite{Feyzmahdavian:13}, whereas Theorem \ref{Theorem 1-2} provides explicit bounds on the decay rate that allow us to quantify the impact of bounded and unbounded time-varying delays on the decay rate.
\end{remark}

%
%

\section{An Illustrative Example}\label{sec:examples}

\begin{figure}[h]
\centering
\includegraphics [width=0.7\columnwidth]{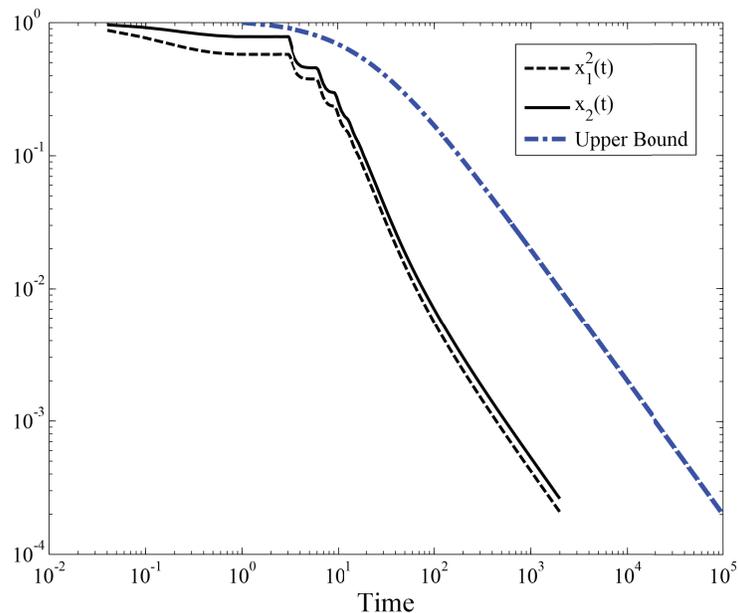}
\caption{Comparison of guaranteed upper bound and actual decay rate of the homogeneous cooperative system (\ref{Example 2}) corresponding to the initial condition $\bphi(t)=(1,1)$, $\forall t\in[-5,0]$.}
\label{Figure 3}
\end{figure}

Consider the continuous-time system (\ref{System 3}) with
\begin{equation}
\label{Example 2}
\f(x_1,x_2)=\begin{bmatrix} -5x_1^3+2x_1x_2\\[0.05cm] x_1^2x_2-4x_2^2 \end{bmatrix},\quad \g(x_1,x_2)=\begin{bmatrix}x_1x_2\\[0.05cm] 2x_1^4 \end{bmatrix}.
\end{equation}
Both $\f$ and $\g$ are homogeneous of degree $p=2$ with respect to the dilation map $\delta^{\r}_{\lambda}(\x)$ with $\r=(1,2)$. Moreover, $\f$ is cooperative and $\g$ is non-decreasing on $\mathbb{R}^2_+$. Since $\f(1,1)+\g(1,1)<\bnull$,  it follows from Theorem \ref{Theorem 3} that the homogeneous cooperative system (\ref{Example 2}) is globally asymptotically stable for any time delays satisfying Assumption \ref{Assumption 5}. Now, consider the specific time-varying delay $\tau(t)=4+\sin(t)$, $t\geq 0$. As $\tau(t)\leq \tau_{\max}=5$ for all $t\geq 0$, Corollary \ref{Corollary 3} can help us to calculate an upper bound on the decay rate of (\ref{Example 2}). Using $\v=(1,1)$ and $r_{\max}=2$, the solutions to (\ref{Corollary 3-1-1}) are $\theta_1=4$, $\theta_2=1$, which implies that
\begin{equation*}
\theta\cong \min\bigr\{\frac{1}{5},\min\{4,1\}\bigl\}=\frac{1}{5}.
\end{equation*}
Thus, the solution $\x(t)$ of (\ref{Example 2}) satisfies
\begin{equation*}
\max\{x_1^2(t),x_2(t)\}=O\left(\frac{1}{\frac{1}{5} t+1}\right),\quad t\geq 0.
\end{equation*}
Figure \ref{Figure 3} gives the simulation results of the actual decay rate of the homogeneous cooperative system~(\ref{Example 2}) and the guaranteed decay rate we calculated, when the initial condition is $\bphi(t)=(1,1)$, for all $t\in[-5,0]$.
%
%
%

\section{Conclusions}\label{sec:conclusions}

This paper has been concerned with delay-independent stability of a significant class of nonlinear (continuous- and discrete-time) positive systems with time-varying delays. We derived a set of necessary and sufficient conditions for global asymptotic stability of continuous-time homogeneous cooperative systems of arbitrary degree and discrete-time homogeneous non-decreasing systems of degree zero with bounded and unbounded time-varying delays. These results show that the asymptotic stability of such systems is independent of the magnitude and variation of the time delays. However, we also observed that the decay rates of these systems depend on how fast the delays can grow large. We developed two theorems for global $\mu$-stability of positive systems that quantify the convergence rates for various classes of time delays. For discrete-time homogeneous non-decreasing systems of degree greater than zero, we demonstrated that the origin is locally asymptotically stable under global asymptotic stability conditions that we derived.

%
%

\bibliographystyle{IEEEtran}
\bibliography{bibliografia}

\end{document}